\documentclass[10pt,journal,compsoc]{IEEEtran}

\newtheorem{definition}{Definition}
\newtheorem{strategy}{Strategy}    

\newtheorem{property}{Property}    

\usepackage{multirow}
\usepackage{amsmath}

\ifCLASSOPTIONcompsoc
  \usepackage[nocompress]{cite}
\else
  \usepackage{cite}
\fi

\ifCLASSINFOpdf
   \usepackage[pdftex]{graphicx}

\else
\fi

\usepackage[ruled,linesnumbered]{algorithm2e}
\usepackage{array}

  \usepackage[caption=false,font=footnotesize,labelfont=sf,textfont=sf]{subfig}
  \usepackage[caption=false,font=footnotesize]{subfig}

\newcommand\MYhyperrefoptions{bookmarks=true,bookmarksnumbered=true,
	pdfpagemode={UseOutlines},plainpages=false,pdfpagelabels=true,
	colorlinks=true,linkcolor={blue},citecolor={blue},urlcolor={blue},
	pdftitle={Anomaly Rule Detection in Sequence Data},
	pdfsubject={Typesetting},
	pdfauthor={Wensheng Gan},
	pdfkeywords={Anomaly detection, sequence,  utility mining, sequential rule.}}
\ifCLASSINFOpdf
\usepackage[\MYhyperrefoptions,pdftex]{hyperref}
\else
\usepackage[\MYhyperrefoptions,breaklinks=true,dvips]{hyperref}
\usepackage{breakurl}
\fi

\usepackage{url}

\begin{document}
\title{Anomaly Rule Detection in Sequence Data}

\author{Wensheng Gan,~\IEEEmembership{Member,~IEEE}, Lili Chen, Shicheng Wan,\\
Jiahui Chen,~\IEEEmembership{Member,~IEEE,}
and Chien-Ming Chen,~\IEEEmembership{Senior Member,~IEEE}

	\IEEEcompsocitemizethanks{\IEEEcompsocthanksitem Wensheng Gan is with the College of Cyber Security, Jinan University, Guangzhou 510632, China. (E-mail: wsgan001@gmail.com)

	\IEEEcompsocthanksitem Lili Chen and Chien-Ming Chen are with the College of Computer Science and Technology, Shandong University of Science and Technology, Qingdao 266590, China. (E-mail: lilichien3@gmail.com and chienmingchen@ieee.org) 

	\IEEEcompsocthanksitem Shicheng Wan and Jiahui Chen are with the School of Computers, Guangdong University of Technology, Guangzhou 510006, China. (E-mail: scwan1998@gmail.com and csjhchen@gmail.com)} %
	
	\thanks{Corresponding author: Chien-Ming Chen}
}

\IEEEtitleabstractindextext{%
\begin{abstract}
	
Analyzing sequence data usually leads to the discovery of interesting patterns and then anomaly detection. In recent years, numerous frameworks and methods have been proposed to discover interesting patterns in sequence data as well as detect anomalous behavior. However, existing algorithms mainly focus on frequency-driven analytic, and they are challenging to be applied in real-world settings. In this work, we present a new anomaly detection framework called DUOS that enables Discovery of Utility-aware Outlier Sequential rules from a set of sequences. In this pattern-based anomaly detection algorithm, we incorporate both the anomalousness and utility of a group, and then introduce the concept of utility-aware outlier sequential rule (UOSR). We show that this is a more meaningful way for detecting anomalies. Besides, we propose some efficient pruning strategies w.r.t. upper bounds for mining UOSR, as well as the outlier detection.  An extensive experimental study conducted on several real-world datasets shows that the proposed DUOS algorithm has a better effectiveness and efficiency. Finally, DUOS outperforms the baseline algorithm and has a suitable scalability.

\end{abstract}

}

\maketitle

\IEEEdisplaynontitleabstractindextext

\IEEEpeerreviewmaketitle

\section{Introduction} \label{sec:introduction}

\IEEEPARstart{W}{ith} a wide range of emerging real-world applications, large amounts of data have been generated. How to effectively perform data analytic in emerging data-driven applications is interesting and challenging. Among this scenario, it has led to a renewed attention in anomaly detection \cite{gupta2013outlier,liu2020event} and security issues \cite{ye2017survey} by leveraging the power of machine learning and data analysis. For example, risk modeling, utility mining \cite{gan2021survey,zhang2021shelf}, intrusion/malware detection \cite{ye2017survey}, and usage behavior anomaly detection \cite{dokas2002data} are indispensable in many fields like cyber-security, market basket analysis, fintech, healthcare, public security and AI safety. In theory and foundation of anomaly detection, the explaining of anomalies/outliers is a fundamental issue \cite{gupta2013outlier}. And outlier pattern detection takes an important role in this research field, such as \cite{fan2016malicious}.

To further motivate outlier pattern detection, in this section, we describe two real-world applications: 

1) \textit{Network intrusion detection}. It has become increasingly important to find the intrusion from a wide variety of cyber threats. However, the standard frequent pattern mining techniques are not helpful to detect emerging cyber threats since most of the cyber threats are unusual. On the network, the number of intrusions is usually a very small fraction of the total network traffic \cite{gupta2013outlier}. Based on the KDD Cup'99 dataset \cite{dokas2002data}, it has demonstrated that the rare class (aka outlier) prediction techniques are much more efficient for detecting intrusive behavior than traditional classification methods.

2) \textit{Targeted marketing }\cite{zadrozny2001learning}. Despite the enormous amount of data, particular interesting goods/events are still quite rare. The so-called rare events are events that occur very infrequently. In targeted marketing, response is typically rare but can be profitable (w.r.t. high-utility) \cite{lazarevic2004data}.  Moreover, a targeted-marketing approach usually lies in the fact the events are sequential in nature. In general, sequence data is commonly seen in targeted marketing application. These rare profitable events can be said as profitable outliers in sequence data. Therefore, utility but not frequency is more appropriate than other factors for evaluating methods for rare event detection in decision making.  

Many techniques have been developed for outlier detection (or called anomaly detection) \cite{gupta2013outlier}. These studies can be broadly classified into two categories: (i) rule-based techniques, and (ii) various data-driven approaches. Note that standard approaches, e.g., frequent pattern mining \cite{agrawal1994fast}, do not work well for rare pattern (outlier) analysis. To effectively identify the domain expert's interests, several measures are commonly used in anomaly detection models, such as frequency (statistically rare point), utility (high-utility value), and risk. There exists a considerable number algorithms of pattern mining for anomaly detection (or in short, PM4AD) \cite{feremans2019pattern,agrawal2015survey}. Up until now, there are many pattern-based anomaly detection algorithms, and most of them studied the issues of pattern representation, evaluation metrics of normal pattern, pattern mining, and computation of the statistical anomaly score.

Sequential data has additional temporal information compared to transactional data. In most of application scenarios, the sequential relations among objects usually have been hidden but more significant information can reveal personalized behaviors \cite{fan2016malicious}. For example, Zhu \textit{et al.} \cite{zhu2016mining} introduced rare sequential topic patterns and then extracted topics in document stream. Identification of hot regions in protein-protein interactions \cite{hsu2007identification} is also related to sequential pattern mining. Therefore, it is critical to capture the underlying information in sequence data. In this paper, we will concentrate on the basic problem of PM4AD in sequence data. Moreover, the discovered rare (infrequent) patterns may have rich underlying information (e.g., sequential information) in many application scenarios, such as economic case investigation, financial internal audit \cite{zhu2016mining}, credit risk assessment \cite{zhu2016mining}, and sequence outlier pattern detection in Internet of Things (IoT) \cite{cao2019efficient}.

\textbf{State-of-the-art and challenges}. In these pattern-based anomaly detection techniques, their computational complexity in detecting rare occurrences of events is one of the key challenges. For example, the rare graph mining technique may easily suffer from costly sub-graph isomorphism \cite{noble2003graph,nandi2016anomaly}. Another challenge comes to that PM4AD in sequence data could generate an enormous number of candidates during the mining and test processes \cite{fan2016malicious}. This problem is exacerbated when the processed data is large scale or the minimum thresholds are set low. In summarize, they may suffer from two intrinsic problems: (i) a high time-complexity for computing rarity and anomaly score, i.e., this PM4AD problem is NP-hard; and (ii) the difficulty in identification of normal pattern, i.e., it is difficult to select suitable metrics to well evaluate normal patterns. In order to incorporate utility into data mining, there are many utility-driven data mining algorithms have been studied \cite{gan2021survey,zida2015efficient}. The total utility of a pattern (e.g., itemset, rule, sequence) is an estimator of the true value which has more useful information than that of the frequency value. Therefore, it is critical for us to capture the outlier patterns that are high-utility anomalies. As a result, how to distinguish the high-utility anomalies instead of those low-utility statistical outliers by incorporating utility is studied in this paper.

\textbf{Major contributions}. Our current work is motivated by the need to detect unusual high-utility sequential rules from a set of sequence data. Designing an effective and efficient outlier pattern detection approach is challenging in nature, for several reasons. First, we need to detect the rare utility-driven patterns. Thus, there is an urgent requirement for PM4AD algorithms that can process massive sequence data efficiently and provide acceptable mining performance. We take into account the fact that anomalous behavior relies on sequence data and it may have a high-utility value. To the best of our knowledge, until now there has been no work in utility-driven rare sequential rule mining for anomaly detection. The primary contributions of this paper can be summarized as follows: 

\begin{itemize}
	
	\item We incorporate utility into outlier detection in sequence data. This work first introduces the concept of utility-aware outlier sequential rule (UOSR), and then formulates the problem of UOSR detection that is more practical for the task of anomaly pattern detection in sequence data.
	
	\item We solve this problem using a novel algorithm named Discovery of Utility-aware Outlier Sequential rules (DUOS) with both frequency and utility measures, as well as sequence weighting factor. A new data structure, \textit{Rule Count Matrix}, is designed to maintain rich information. And the designed utility table can avoid repeatedly scanning the database and then reduce the execution time. 
	
	\item For efficiency improvement, we developed several optimization techniques that utilize upper bounds on utility and rare rule. Based on the anti-monotonous properties of upper bounds, several pruning strategies were designed to accelerate the mining process. 
	
	\item Experimental results demonstrate that the proposed DUOS algorithm is capable for effectively detecting rare high-utility anomalies in real sequence datasets. 
\end{itemize}

The remainder of this paper is organized as follows. In Section \ref{sec:relatedwork}, related works are briefly reviewed. Section \ref{sec:preliminary} introduces essential definitions and formulates the problem of UOSR detection. The details of proposed DUOS method with several upper bounds and compact data structures, are described in Section \ref{sec:algorithm}. Extensive experiments are conducted and the results are presented in detail in Section \ref{sec:experiments}. Finally, the conclusion and future work are drawn in Section \ref{sec:conclusion}.

\section{Related work}  \label{sec:relatedwork}

This section introduces some related work of frequency-based pattern mining, and pattern mining for anomaly detection (PM4AD). Outlier detection techniques have been extensively studied and successfully applied in different domains. Note that this paper mainly focuses on the studies of PM4AD, especially in sequence data.

\subsection{Pattern Mining}

Pattern mining is one of the key processes of data mining and knowledge discovering. In the field of pattern mining, frequent itemset mining (FIM) \cite{han2004mining, zaki2000scalable} and sequential pattern mining (SPM) \cite{srikant1996mining,zaki2001spade,tran2016mining} are the two common tasks to automatically determine whether a pattern (e.g., itemset, rule, sequence) is frequent or not. Note that a pattern in FIM or SPM is purely based on the number of its occurrences. In the past decades, FIM has been extensively studied, such as Apriori \cite{agrawal1994fast} and FP-growth \cite{han2004mining}. While SPM aims at finding those frequently occurring sequential patterns in a sequence database which is different from transaction database. The records in a sequence database have the key characteristic information w.r.t. timestamp, and they often contain a set of subsequences of events that usually have a chronological order. There are many real-life sequence data \cite{fournier2017survey,gan2019survey}, for example, the customer shopping records, biological sequences, and video session records.  Up until now, a lot of SPM algorithms have been proposed, such as GSP \cite{srikant1996mining}, SPADE \cite{zaki2001spade}, and PrefixSpan \cite{pei2004mining}, MSPIC-DBV \cite{van2018mining}. The well-known PrefixSpan adopts a prefix-projection method to discover sequential patterns. In addition, sequential rule mining (SRM) \cite{fournier2012cmrules,fournier2014erminer} is different from SPM although both of them aim at dealing with symbolic sequence datasets. The discovered rules in SRM not only have rich underlying sequential information, but also have a high confidence. Recently, utility-driven SPM \cite{gan2020fast,zhang2021tkus,zhang2021shelf} and utility-driven SRM \cite{zida2015efficient} also have been extensively studied.

The fundamental problem of most pattern mining algorithms is the lack of data associated with rare cases. For example, rare patterns (e.g., itemsets, rules) tend to cover only a few records/instances. Obviously, the patterns that rarely occur in very few records will be normally pruned by pattern mining approaches. However, these infrequent patterns warrant special attention since they usually mean the anomaly and interesting. How to cater for these specific types of rare/anomaly pattern detection is a difficult task, especially when we consider the utility factor instead of frequency. To summary, these typical approaches can discover different kinds of frequent patterns, but they do not adequately capture outlier patterns.

\subsection{Pattern Mining for Anomaly Detection}

There exists a considerable number algorithms of pattern mining for anomaly detection (PM4AD) \cite{feremans2019pattern,agrawal2015survey}. Up until now, there are many existing pattern-based anomaly detection algorithms that focusing on a particular combination of rare pattern representation, pattern mining, and computation of the anomaly score. In general, there are many kinds of rare pattern, such as rare association rule, rare itemset, rare sequential rule, and details can be referred to \cite{koh2016unsupervised}. In the past decades, pattern-based anomaly detection has been widely studied. Wong \textit{et al.} \cite{wong2002rule} proposed a rule-based anomaly pattern detection for detecting disease outbreaks. Graph-based anomaly detection was also studied in \cite{noble2003graph}. Nandi \textit{et al.} \cite{nandi2016anomaly} utilized program control flow graph to address the problem of anomaly detection from execution logs. In the field of anomaly detection on trajectory data, a framework called MT-MAD \cite{lei2016framework} was proposed to find anomalous movement behavior. In these frameworks and algorithms, the processed data usually consists mostly of normal records, along with a very small percentage of anomalous records. Recently, there are some deep learning-based methods that designed for anomaly detection, as reviewed in \cite{chalapathy2019deep,pang2021deep}. How to efficiently discover the anomaly behavior from large-scale dataset is still challenging.

Although pattern-based anomaly detection have been applied to the problem of discovery anomalies. In the literature, little attention considers the external constraints (e.g., utility, risk, unit price) \cite{geng2006interestingness} for discovering abnormal patterns, even for reducing the runtime and memory cost of a PM4AD algorithm. In the set of discovered patterns, their frequency is one of the objective statistical properties, while their utility is a subjective measure \cite{gan2021survey}. In other words, a subjective measure is mainly depending on the prior knowledge from human and different from the objective one. In the past, as an alternative interestingness measure, utility has been adopted in many data mining algorithms to discover useful patterns and knowledge \cite{gan2021survey,gan2018survey}. This motivates us to incorporate utility-aware measure into anomaly detection in sequence data.

\subsection{Outlier Detection in Sequence Data} 

As mentioned previously, the occurred events (data) are usually sequential in nature. There are a wide range of emerging real-world applications that focus on dealing with sequence data \cite{rasheed2013framework} and stream data \cite{na2018dilof}. For example, Fan \textit{et al.} \cite{fan2016malicious} studied the problem of malicious sequential pattern mining for automatic malware detection. There are also several methods that detect outliers from symbolic sequence databases \cite{budalakoti2008anomaly,hofmeyr1998intrusion,warrender1999detecting}. In these studies, the outliers are defined as input sequences, and the symbolic sequence is different from any other sequence based on distance (i.e., edit distance) measures.   Some methods \cite{keogh2005hot,keogh2007finding} studied the issue of identifying anomalous subsequences within a single long sequence.  And some methods identify the interesting patterns in a sequence in which their frequency of occurrences is abnormal \cite{keogh2002finding}. However, these algorithms focus on time series composed of numerical values. In other words, they do not deal with symbolic sequence data, as surveyed in \cite{chandola2010anomaly}. Moreover, these techniques cannot solve the addressed problem of detecting utility-aware outlier sequential rules that violate the expected behaviors. Notice that several anomaly detection problems require the user to specify candidate outlier patterns \cite{keogh2002finding}. They determine if a query pattern $\alpha$ in a long test sequence $t$ is anomalous, depending on the frequency of occurrence of $\alpha$. Lu \textit{et al.} \cite{lu2017unsupervised} studied this task with a deep architecture.  Obviously, these methods are not effective in discovering  outlier sequential patterns with various constraints, e.g., utility.  In our work, we focus on automatically detecting all outlier sequential rules without requiring assumed outlier patterns. Besides, an assumed normal training set is not given beforehand. This will lead to proposing utility-based techniques that are capable of detecting sequential outliers without many assumptions.

\section{Preliminary and Problem Statement} 
\label{sec:preliminary}

In this section, we first introduce some definitions and properties that are related to the DUOS algorithm. Then we formulate the problem of utility-aware outlier sequential rule detection.

\subsection{Preliminary}

Note that \textit{I} = \{\textit{i}$_{1}$, \textit{i}$_{2}$, $\ldots$, \textit{i$_{l}$}\} is a collection of items. Each item $i$ is attached to an external positive value indicating the unit profit or weight of the item, which is referred to as external utility and denoted as $p(i)$.  An itemset \textit{I}$_x$ is a subset of \textit{I}, consisting of several distinct items. And a sequence \textit{S} is consisted of groups of  itemsets, without loss of generality, arranged in lexicographical order and expressed as $\prec$. In addition, each item \textit{i} in a sequence \textit{S} also corresponds to a positive purchase quantity called internal utility and expressed as $q(i, S)$.

A sequence database \textit{SD} is composed of a few sequences that can be represented as \{\textit{S}$_{1}$, \textit{S}$_{2}$, $\ldots$, \textit{S$_{n}$}\}. It is required to declare that there cannot exist two identical items in a sequence. For ease of explanation, a sequence database and a profit table are shown in Table \ref{table:db} and Table \ref{table:profit}, respectively. There are four sequences in Table \ref{table:db} and each letter represents an item. Items and their quantities are in parentheses, and items between curly brackets indicate an itemset. The brackets are eliminated in case  there is merely a single item in an itemset. Since items in an itemset are not in chronological order, $a$ and $b$ occurs simultaneously in $S_1$, followed by $c$, then $f$, $g$, and $e$ in succession.

\begin{table}[!htbp]
	\centering
	\small
	\caption{Sequence database.}
	\label{table:db}
	\begin{tabular}{|c|c|c|c|}
		\hline
		\textbf{\textit{SID}} & \textbf{Sequence (item, quantity)} & \textbf{\textit{SEU}} \\ \hline \hline
		$ S_{1} $ & 	\{(\textit{a}, 1) (\textit{b}, 2)\}  (\textit{c}, 2) (\textit{f}, 3) (\textit{g}, 2)  (\textit{e}, 1) &  \$27 \\ \hline
		
		$ S_{2} $ & 	\{(\textit{a}, 1) (\textit{d}, 3)\} (\textit{c}, 4) (\textit{b}, 2) \{(\textit{e}, 1) (\textit{g}, 2)\}  & \$40 \\ \hline
		
		$ S_{3} $ &	    (\textit{a}, 1) (\textit{b}, 2) (\textit{f}, 3) (\textit{e}, 1)
		&  \$15 \\ \hline
		
		$ S_{4} $ &	   	\{(\textit{a}, 3) (\textit{b}, 2) (\textit{c}, 1)\}	\{(\textit{f}, 1) (\textit{g}, 3)\}  &  \$16 \\ \hline
		
	\end{tabular}
\end{table}

\begin{table}[!htbp]
	\centering
	\small
	\caption{Unit utility of each item}
	\label{table:profit}
	\begin{tabular}{|c|c|c|c|c|c|c|c|}
		\hline
		\textbf{Item}    &  $ a $  &  $ b $  &  $ c $   & $ d $  & $ e $ & $ f $& $ g $ \\ \hline
		\textbf{Utility (\$)} &  $ 1 $  &  $ 2 $  &  $ 5 $  & $ 4 $  & $ 1 $ & $ 3 $ & $ 1 $ \\ \hline
	\end{tabular}
\end{table}

\begin{definition}
	\label{def_n1}
	\rm We define a sequential rule $X \Rightarrow Y$ as a relationship between $X$ and $Y$ where non-empty itemset $X$ and $Y$ belong to $I$ and they do not intersect with each other. This relationship can be illustrated by the statement that when $X$ occurs in a sequence, then $Y$ will subsequently takes place \cite{zida2015efficient}. It is important to notice that the concept of sequential rule is different from association rule \cite{han2004mining, zaki2000scalable} or sequential pattern \cite{srikant1996mining,zaki2001spade}.
\end{definition}

\begin{definition}
	\label{def_n2}
	\rm For a rule $r$ : $X \Rightarrow Y$, we consider it occurring in a sequence $S_c$ = \{$I_1$, $I_2$, \dots, $I_m$\} under the circumstance that there is an integer $p$ such that $ m> p \geq$ 1, $X \subseteq \sum_{i = 1}^{p} I_i$ and $Y \subseteq \sum_{i = p + 1}^{m} I_i$
\end{definition}

\begin{definition}
	\label{def_n3}
	\rm	Assuming $r$: $X \Rightarrow Y$ consists of two itemset $X$ and $Y$ where $|X|$ = $k$, $|Y|$ = $m$, the size of $r$ is defined as $k \ast m$.  Moreover, there exists another rule $r'$ that has the size of $g \ast h$, we believe that the size of $r$ is greater than that of $r'$ if $k \geq g$ and $m > h$, or $k > g$ and $m \geq h$.  
\end{definition}

For example, a rule $r$: \{$a$, $b$\} $\Rightarrow$ \{$e$, $g$\} occurs in sequence $S_1$ and $S_2$ in the given running example. Obviously, the size of $r$ is 2 $\ast$ 2. In addition, the size of $r'$: \{$a$\} $\Rightarrow$ \{$e$, $g$\} is 1 $\ast$ 2. Therefore, $r$ is greater than $r'$ naturally.

In this paper, we primarily employ the pattern growth approach to acquire the complete set of sequence rules. Therefore, initially the size of rules is 1 $\ast$ 1. In order to extract all the rules, it is wise to expand rules.

\begin{definition}
	\label{def_n4}
	\rm Given a rule $r$: $X \Rightarrow Y$, an item $i$ is expanded to the left of $r$, i.e., $X \cup i \Rightarrow Y$, which is called \textit{left expansion}. Similarly, expand an item $i$ to the right of the rule, i.e., $X \Rightarrow Y  \cup i$, which is referred as \textit{right expansion}. 
\end{definition}

\begin{definition}
	\label{def_n5}
	\rm The identifier of sequences containing itemset $X$ are preserved in a collection designated as $sids(X)$. Similarly, the set of sequences containing rule $r$ is kept in $sids(r)$. In addition, we define the support of a rule $r$: $X \Rightarrow Y$ as the number of sequences comprising $r$ divided by the length of the given database, and formally denote it as $sup(r)$ = $|sids(r)|$/$|SD|$. Similarly, we define the confidence of a rule  $r$ as the support of $r$ divided by the support of the antecedent of $r$, i.e., $X$, and express it as $conf(r)$ = $|sids(r)|$/$|sids(X)|$.
\end{definition}

For example, it is evident that the rule $r$: $a \Rightarrow b$ appears in sequence $S_2$ and $S_3$ , and $a$ exists in four sequences with reference to Table \ref{table:db}. Therefore,  it is fairly straightforward to calculate that $sup(r)$ = $|sids(r)|$/$|SD|$ = 2/4 = 0.5, and $conf(r)$ = $|sids(r)|$/$|sids(a)|$ = 2/4 = 0.5.

\begin{definition}
	\label{def_n6}
	\rm Suppose there exists an item $i$ and an itemset $X$ in a sequence $S_c$, the utility of $i$ in this sequence can be formulated as $u(i, S_c)$ = $q(i, S_c) \times p(i)$. Furthermore, the utility of $X$ in $S_c$ is denoted as $u(X, S_c)$ = $\sum_{i \in X \wedge X \in S_c}u(i, S_c)$. We denote the utility of a rule $r$ \cite{zida2015efficient}: $X \Rightarrow Y$ in a sequence $S_c$ as $u(r, S_c)$ = $u(X \cup Y, S_c)$. And the utility of $r$ in a sequence database is denoted as $u(r)$ = $\sum_{S_c \in SD}u(r, S_c)$ where $r$ appears in $S_c$.
\end{definition}

Consider the example given above, the utility of $a$ and $b$ in sequence $S_2$ can be calculated as \$1 and \$4, respectively. Then, the utility of $r$  in $S_2$ equals to \$1 + \$4 = \$5. Hence, the utility of $r$ in this sequence database is $u(r)$ = $u(r, S_2)$ + $u(r, S_3)$ = \$10. A rule is regarded as \textit{frequent} under the circumstance that its support is no less than \textit{maxsup}, which is a threshold to distinguish whether it is frequent or rare. A rule, whose support does not exceed a given threshold \textit{maxsup} and is not less than a given value \textit{minsup} (a minimum support threshold of a requested rule), would be considered as \textit{infrequent} (or called rare).

\begin{definition}
	\label{def_n10}
	\rm A rare high-utility sequential rule (RHUSR) is required to satisfy that  \textit{minsup} $ \leq sup(r) < $ \textit{maxsup}, $u(r) \geq $ \textit{minutil} (the minimum utility threshold) and meanwhile $conf(r) \geq $ \textit{minconf} (the minimum confidence threshold) where  0 $\leq $ \textit{minsup} $ \le $ \textit{maxsup} $\leq$ 1.0, \textit{minutil} $ \in R^{+}$, and \textit{minconf} $ \in$ [0, 1.0] are established according to various scenarios requirements. Otherwise, it is considered as unqualified.
\end{definition}

Obviously, the RHUSRs is a kind of rare patterns and it also has a high-utility value and a high confidence. Based on the above definitions, all the RHUSRs in Table \ref{table:db} are listed in the following Table \ref{table:result}  when \textit{minsup} is set as 0.25, \textit{maxsup} is set as 1,  \textit{minutil} is \$41, and \textit{minconf} is 0.7.

\begin{table}[!htbp]
	\centering
	\small
	\caption{Rare high-utility sequential rules}
	\label{table:result}
	\begin{tabular}{|c|c|c|c|}
		\hline
		\textbf{Rule}  &  \textbf{Support}   &  \textbf{Utility (\$)}  &  \textbf{Confidence}  \\ \hline
		$\{a, b, c\} \Rightarrow \{g\} $ & 0.75 & 57.0 & 1.0 \\ \hline
		$\{a, c\} \Rightarrow \{g\} $  & 0.75 & 45.0 & 1.0 \\ \hline
		$\{b, c\} \Rightarrow \{g\} $  & 0.75 & 52.0 & 1.0 \\ \hline
	\end{tabular}
\end{table}

\subsection{Problem Statement}
\label{Problem statement}

Intuitively, a sequential rule will be frequent  if it occurs frequently across multi-sequences (w.r.t. \textit{SD}) and this is independent of any of its super-patterns. The high-utility outliers capture the rules that occur rarely and violate the typical sequential rules in the processed sequence data as formally defined next.

\begin{definition}
	\rm Consider the utility factor, if a sequential rule has the following conditions: 1) it is infrequent/rare, 2) it has a high-utility value no less than the minimum utility threshold, and 3) it has a distinguished anomaly score compared to others, then it is called a utility-aware outlier sequential rule (UOSR). Here the anomaly score of a record w.r.t. UOSR depends on its outlier factor, which will be described in Subsection \ref{section:outlier}.
\end{definition}

\textbf{Problem Statement:} Given a $q$-sequence database that contains both normal and anomalous records, the problem of utility-aware outlier pattern detection in sequence data is to discover the complete set of utility-aware outlier sequential rules (UOSRs) across multi-sequences in the given data, with a set of parameter settings.

So far, various pattern-based methods used to detect anomalies have been discussed in Section  \ref{sec:relatedwork}. The problem in this paper is to identify the high-utility anomalous records among sequence data. We will present a novel utility-aware outlier detection method in the following sections.

\section{The DUOS Algorithm}
\label{sec:algorithm}

\begin{figure}[!htbp]
	\centering
	\includegraphics[scale=0.58,height=0.21\textheight,width=1\columnwidth]{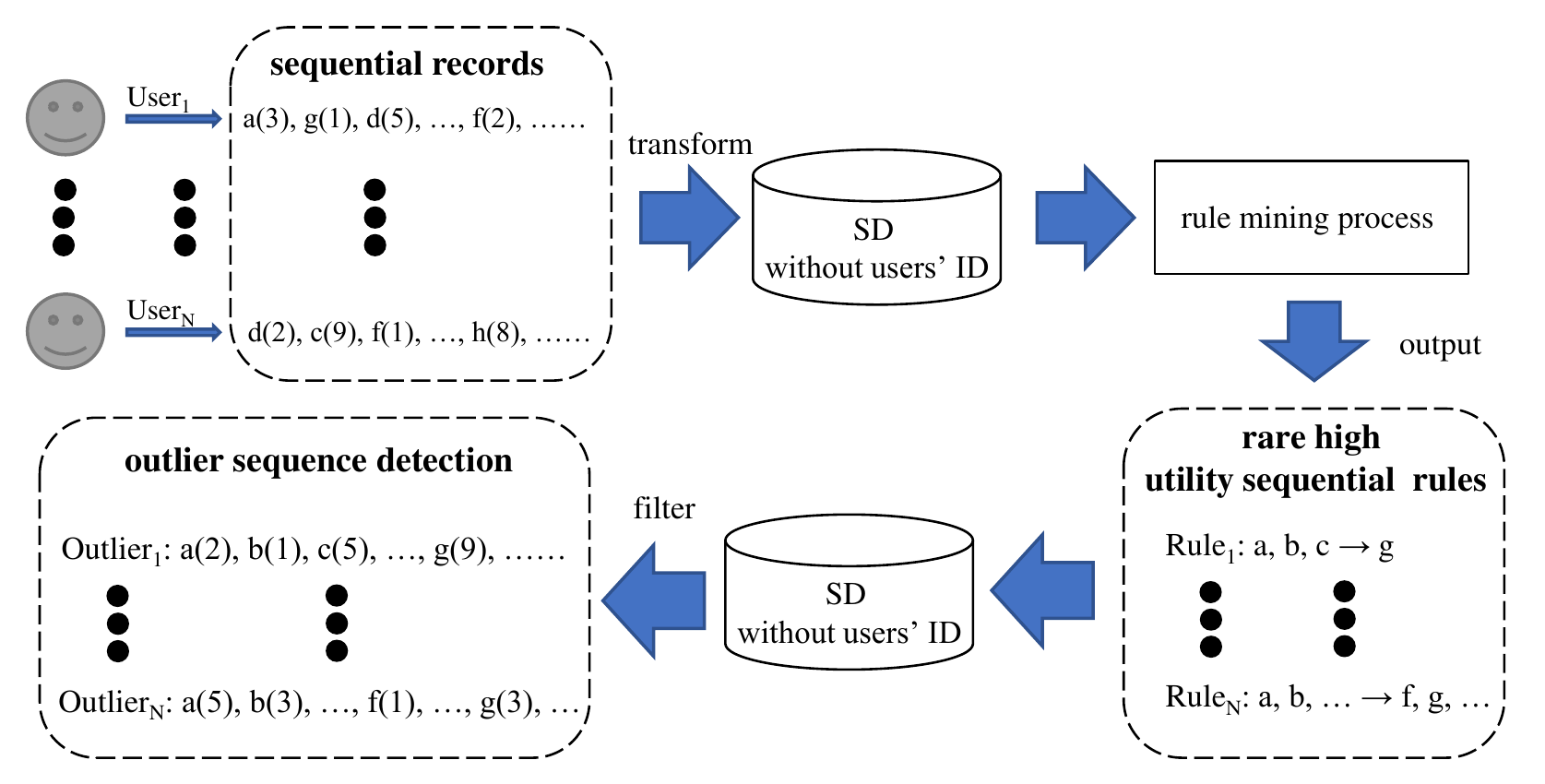}
	\caption{Flowchart of the DUOS framework.}
	\label{fig:Workflow}
\end{figure}

The frequency-based PM4AD methods are more generalizable and intuitive because we can easily compute their co-occurrence and anomaly score. However, the utility-driven method is challenging since the utility measure is neither monotonic nor non-monotonic. Within economics, the concept of utility is used to model worth or value. Thus, there are some complexities and challenges in PM4AD that requires advanced approaches. To tackle these challenges, an optical framework is constructed and demonstrated in Fig. \ref{fig:Workflow}. 

Fig. \ref{fig:Workflow} presents an overview of the proposed DUOS framework for detecting anomalous records. Two primary operations are involved in this framework to identify some outliers, namely rule mining and outlier detection. Rule mining requires the discovering of rare high-utility sequential rules (RHUSRs) with some efficient pruning strategies and structures, and outlier detection is to determine whether the discovered RHUSRs are outliers, depending on the basis of the previously extracted rules. We will explain the details in the following sections.

\subsection{RHUSR Mining}

In this subsection, we first introduce the concept of sequence estimated utility with the downward closure property. Then, the utility table structure is constructed. Thereafter, the algorithm concerning RHUSR mining is elaborately described. Finally, to optimize the proposed approach, several pruning strategies are designed.

\subsubsection{Sequence Estimated Utility}

Rare high-utility sequential rule mining involves the calculation of support, confidence, and utility of rules. Here, we investigate on utility. Notoriously, the utility concept does not hold the downward closure property. Therefore, it is extremely time consuming to calculate the utilities of entire patterns or rules in databases. In order to minimize the probability of computing utilities, we present the sequence estimated utility as an upper bound on utility that can hold the anti-monotonic property.

\begin{definition}
	\rm The sequence utility of a sequence $s$ is defined as the sum of the utilities of all items in $s$, and denoted as $SU(s)$ = $\sum_{i \in s}u(i, s)$. We define the sequence estimated utility of an item $i$ as the summary of the sequence utility of sequences supporting $i$, and denote it as $SEU(i)$ = $\sum_{i \in s \wedge s \in SD}SU(s)$. We define the sequence estimated utility of a sequential rule $r$ as the summary of the sequence utility of sequences that contains $r$, and denote it as $SEU(r)$ = $\sum_{s \in seq(r)}SU(s)$.
\end{definition}

It is obvious that the sequence estimated utility of an item or a rule is necessarily greater than or equal to its true utility. Hence, if the sequence estimated utility of an item or a rule is less than \textit{minutil}, then it is indisputable that its utility is definitely less than \textit{minutil}.

\begin{strategy}
	\label{stra_1}
	\rm If $SEU$ of an item $i$ is less than \textit{minutil}, then any rule that contains this item is not RHUSR for sure and can be directly pruned.
\end{strategy}

\begin{strategy}
	\label{stra_2}
	\rm If $SEU$ of a rule $r$ is less than \textit{minutil}, then this rule as well as its expansions are certainly not a RHUSR and can be directly pruned.
\end{strategy}

\subsubsection{Utility Table Structure}

To obtain all the sequential rules, we mainly adopt the pattern growth approach whose operation is to extend the shorter ones by \textit{left expansions} or \textit{right expansions}.

\begin{definition}
	\rm Given a sequential rule $r$: $X \Rightarrow Y$ which exists in a sequence $s$. An item $i$ requires the condition  that $i \prec j$, $\forall j \in X, i \notin Y$, and simultaneously $X$ $\cup$ $\{i\}$ $\Rightarrow$ $Y$ appears in $s$ should be satisfied for left-expanding $r$.  Let a sequential rule $r$: $X \Rightarrow Y$ exist in a sequence $s$. An item $i$ is able to right-expand $r$ under the circumstance that $i \prec j$, $\forall j \in Y, i \notin X$, and moreover $X \Rightarrow Y$ $\cup$ $\{i\} $ also appears in $s$.
\end{definition}

For convenience, if we preserve the items in a sequence $s$ that can merely left-expand $r$ in a collection \textit{onlyLeft}($r, s)$, the items can only right-expand $r$ in a set \textit{onlyRight}($r, s)$, and the items can extend $r$ by both left-expansion and right-expansion in \textit{leftRight}($r, s)$. However, in practice we have discovered that  a rule can be obtained according to different combinations of left and right expansions. 

A straightforward and effective solution is to prohibit right expansions after left expansions, but allow left expansions after right expansions, and vice versa. Furthermore, another complexity is that a rule can be accessed by left-expanding or right-expanding a different item, which may easily cause rule redundancy. To overcome this challenge, we require that an item should be greater than each item in the antecedent when left-expanded, and meanwhile it should be greater than each item in the consequent when right-expanded. After introducing these basic concepts, we formally describe the definition of the utility table.

\begin{definition}
	\label{def_6}
	\rm The \textit{utility table} of a rule $r$ is a collection of tuples, and each of tuples consists of five columns, namely ($sid$, \textit{iutil}, \textit{lutil}, \textit{rutil}, \textit{lrutil}) \cite{zida2015efficient}. Each row of the tuple corresponds to the critical information of the sequence $s$ containing $r$. The $sid$ is the sequence identifiers of the sequence containing $r$. The  \textit{iutil} is defined as the utility of $r$ in $s$. The \textit{lutil} element in $s$  is designed as the sum of utilities of items in the set of \textit{onlyLeft}($r, s)$. Similarly, the \textit{rutil} element in $s$  is defined as the total of utilities of items in  \textit{onlyRight}($r, s)$, the \textit{lrutil} is the summary of utilities of items in \textit{leftRight}($r, s)$.
\end{definition}

\begin{figure}[!htbp]
	\centering
	\includegraphics[scale=0.5]{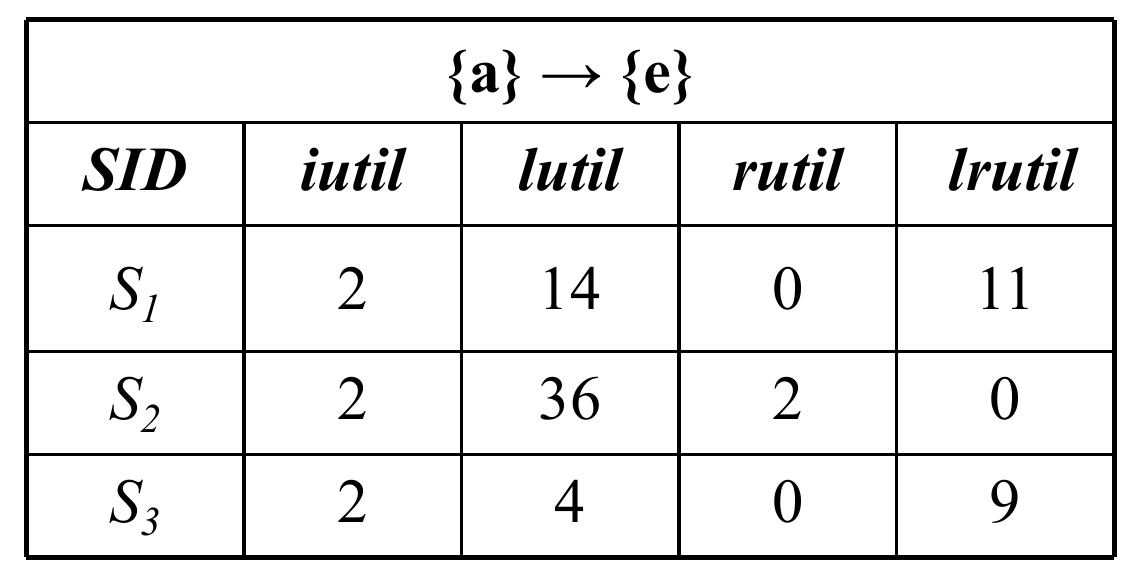}
	\caption{The utility table.}
	\label{fig:utilityTable}
\end{figure}

The utility table of ${a} \Rightarrow {e}$ in the running example is shown in Fig. \ref{fig:utilityTable}. Clearly, the \textit{iutil} is designed to calculate the utility of a sequential rule $r$. Since longer rules are extended by shorter ones, to determine whether a rule can be extended left or right in terms of utility restrictions, we maintain the maximum utility \textit{lutil}, \textit{rutil}, and \textit{lrutil} that can be achieved by the left expansion and right expansion of each rule in each sequence. One great advantage of the utility table structure \cite{zida2015efficient} to maintain information is that it avoids repeatedly scanning the database and can reduce the execution time. The utility, support and confidence of a rule can be obtained from a utility table. We refer to the utility table corresponding to rule $r$ as $UT(r)$. The sum of \textit{iutil} in $UT(r)$ is expressed as $u(r)$. Besides, the support of $r$ equals to the number of rows in $UT(r)$. In addition, some other properties based on utility also can be obtained.

\begin{property}
	\label{pro_1}
	\rm Given a sequential rule $r$ and its corresponding utility table $UT(r)$. The sum of \textit{iutil}, \textit{lutil}, \textit{rutil}, and \textit{lrutil} is the upper bound of utility of $r$ as well as its expansions. Furthermore, it is tighter than $SEU(r)$.
\end{property}

\begin{property}
	\label{pro_2}
	\rm The sum of \textit{iutil}, \textit{lutil}, and \textit{lrutil} is the upper bound of utility of $r$ as well as its left-expansions, which is also tighter than $SEU(r)$.
\end{property}

Since the proposed algorithm employs rule expansion method to extend the initial rules of 1*1 to discover all the desired sequential rules. In order to access the utility table of larger sequential rules without scanning the database, we present the following construction procedure.

\begin{definition}
	\label{def_8}
	\rm Suppose a sequential rule $r$ is expanded left or right into $r'$ with an item $i$. In the sequence $s$ that contains $r$ and $r'$, the tuple of $r$ is ($s$, \textit{iutil}, \textit{lutil}, \textit{rutil}, \textit{lrutil}), while $r'$ is ($s$, \textit{iutil'}, \textit{lutil'}, \textit{rutil'}, \textit{lrutil'}).	The parameters involving in $r'$ can be calculated as follows.	
	\begin{equation}
	\label{equ_1}
		iutil' = iutil + u(i, s). 
	\end{equation}	
	\begin{equation}
	\label{equ_2}
		lutil' = lutil - \sum u(j, s) - u(i, s), 
	\end{equation}
	where $ j \notin$ \textit{onlyLeft}($r', s) \wedge j \in$ \textit{onlyLeft}($r, s)$, $i \in$ \textit{onlyLeft}($r, s)$.
	
	\begin{equation}
	\label{equ_3}
		rutil' = rutil - \sum u(j, s) - u(i, s), 
	\end{equation}
	where $ j \notin $ \textit{onlyRight}($r', s) \wedge j \in $ \textit{onlyRight}($r, s)$, $i \in$ \textit{onlyRight}($r, s)$.
		
	\begin{equation}
	\label{equ_4}
		lrutil' = lrutil - \sum u(j, s) - u(i, s), 
	\end{equation}
	where $ j \notin $ \textit{leftRight}($r', s) \wedge j \in $ \textit{leftRight}($r, s)$, $i \in $ \textit{leftRight}($r, s)$.
\end{definition}

\subsubsection{Optimization}

This section aims at enhancing the performance for mining sequential rules. In view of the three parameters (support, utility, and confidence) that are involved in the rare high-utility sequential rule mining, we optimize them separately.

With respect to support measure, since the claimed rules is rare, the support value is necessary to satisfy the range between \textit{minsup} and \textit{maxsup}. Assuming the support of a rule is larger than \textit{minsup}, this rule may be frequent or rare, and then it is possible for the expansions of this rule to be rare. However, if its support is less than \textit{minsup}, neither this rule nor its expansions would be rare. On the basis of this, we suggest an effective pruning strategy below.

\begin{strategy}
	\label{stra_3}
	\rm In case the support of a rule is less than \textit{minsup}, then this rule and all its super-rules can be directly pruned, without left and right expansions.
\end{strategy}

To further exploit the downward closure property of support (a filter with rarity), we design a matrix structure named \textit{Rule Count Matrix} (RCM)  to hold the support of all promising initial sequential rules. Let a set of sequential rules in the form of $r$: $i \Rightarrow j$, the value of $RCM(i ,j)$ is equivalent to the support of $r$.

Suppose $r$: $X \Rightarrow Y$ and $r'$: $X \Rightarrow Y  \cup \{i\}$, i.e., $r'$ is the right expansion of $r$. If any item $j$ in $X$ such that $RCM(j ,i) <$ \textit{minsup}, then $r'$ is definitely not qualified. In the same way, $r''$: $X \cup \{i\} \Rightarrow Y$ is the left expansion of $r$. If any item $j$ in $Y$ such that $RCM(i ,j) <$ \textit{minsup}, then $r''$ is definitely not qualified. According to this observation, we design the following strategies.

\begin{strategy}
	\label{stra_4}
	\rm Let a sequential rule $r$ has left expand with an item $i$. If any item $j$ in consequent has $RCM(i ,j) <$ \textit{minsup}, then we stop the left expansion regarding $i$.
\end{strategy}

\begin{strategy}
	\label{stra_5}
	\rm Let a sequential rule $r$ has right expand with an item $i$. If any item $i$ in antecedent has $RCM(j ,i) <$ \textit{minsup}, then we stop the right expansion regarding $i$.
\end{strategy}

As for utility, we have already designed two pruning strategies based on the sequence estimated utility. Here we further introduce two new pruning strategies based on Property \ref{pro_1} and Property \ref{pro_2}, as described below.

\begin{strategy}
	\label{stra_6}
	\rm Given a sequential rule $r$ and its utility table $UT(r)$. If the sum of \textit{iutil}, \textit{lutil}, \textit{rutil}, and \textit{lrutil} in $UT(r)$ is less than \textit{minutil}, then the rule and any of its right expansions can be pruned immediately.
\end{strategy}

\begin{strategy}
	\label{stra_7}
	\rm Suppose a sequential rule $r$ holds a utility table $UT(r)$. If the sum of \textit{iutil}, \textit{lutil}, and \textit{lrutil} in $UT(r)$ is less than \textit{minutil}, then the rule and any of its left expansions can be pruned directly.
\end{strategy}

To calculate the confidence of a rule, it is necessary to know its support in a given database and the support of the antecedent of the rule. Calculating the support of a rule in the utility table is easy, however, it is hard to find the support of its antecedent. A simple and straightforward way is to iterate the database to obtain the support, which is extremely time consuming and inefficient. To address this problem, a bit vector \cite{tran2016mining} for each item is created. In our setting, the \textit{p}-th bit of the bit vector indicates whether the item appears in the \textit{p}-th sequence or not, if it does, then the \textit{p}-th bit is set to 1, otherwise it is 0. For example, since $a$ appears in each sequence and its bit vector is 1111, while $e$ only occurs in the first three sequences, whose bit vector is 1110. The way to determine whether a pattern appears in a sequence is to take the intersection of all items in the pattern. Then the bit vector of $\{ae\}$ is the intersection of the bit vectors of $a$ and $e$, which is 1110 and the support of $\{ae\}$ is 3. The advantages of this approach are that the bit vector operation is very fast and it occupies very little memory.

To summary, the designed strategies \ref{stra_3}, \ref{stra_4}, and \ref{stra_5} can prune unpromising sequential rules from support perspective, while the strategies \ref{stra_1}, \ref{stra_2}, \ref{stra_6}, and \ref{stra_7} are depended on utility, and the bit vector optimization utilizes the confidence aspect. The order of applying these strategies is shown later in the introduction of the proposed algorithm. These pruning strategies can be partially applied to the algorithm, nevertheless, a better performance is obtained by all applications.

\subsection{Rule-based Outlier Detection} \label{section:outlier}

So far the RHUSRs have been mined, next we should exploit those sequences of outliers that are anomalous based on the discovered rules. It is important to note that the more rare high-utility sequential rules are existed in a sequence, the higher probability that the sequence becomes an outlier. Therefore, we propose a sequence weighting factor (SWF) to measure the attributes.

\begin{definition}
	\label{def_9}
	\rm Let \textit{RHUSR} be the collection of complete rare high-utility sequential rules in the database \textit{SD}, and \textit{maxsup} be the maximum support threshold. For each subsequence $s$, \textit{SWF}($s$) is defined as the number of RHUSRs in a sequence divided by the size of the set \textit{RHUSR}. For each rule $r$ in \textit{RHUSR}, the deviation factor (DF) of $r$ is defined as: $DF(r)$ = (\textit{maxsup} - $sup(r)$)/$|$\textit{RHUSR}$|$.
\end{definition}

For the evaluation of the degree of deviation for a special rule, the following formula of utility-aware outlier sequential rule (abbreviated as UOSR) can be referred.

\begin{definition}
	\label{def_11}
	\rm  For each subsequence $s$ in a sequential rule $r$, an outlier factor (OF) is defined as $OF$($s$) = 1 - (\textit{SWF}($s$)*$\sum_{r \in RHUSR  \wedge  r \subseteq s}DF$($r$)). Given a user-specified outlier threshold $v$, if $OF(s) \geq v$, we say that $s$ is a utility-aware outlier sequential rule (UOSR).
\end{definition}


\begin{algorithm}[h]
	\label{OutlierDetection}
	\caption{OutlierDetection(\textit{RHUSRs}, $v$, \textit{maxsup})}
	\LinesNumbered 
	\KwIn{\textit{RHUSRs}, the set of rare high-utility sequential rules; $v$, the outlier threshold; \textit{maxsup}, the maximum support threshold.} 
	\KwOut{a set of \textit{UOSRs}.} 
	\For{each rule $r \in$ \textit{RHUSRs}}{
		$DF(r)$ = (\textit{maxsup} - $sup(r)$)/$|$\textit{RHUSRs}$|$\;
	}
	\For {each sequence $s$ in \textit{SD}}{
		$S(s)$ = 0, $A(s)$ = 0\;
		\For {each rule  $r \in $ \textit{RHUSRs}}{
			\If{$s$ contains $r$}{
				$S(s)$ = $S(s)$ + 1\;
				$A(s)$ = $A(s)$ + $DF(r)$\;
			}
		}
		
		\textit{SWF}($s$) = $S(s)$/$|$\textit{RHUSRs}$|$\;
		\textit{OF}($s$) = 1 - \textit{SWF}$(s) \times A(s)$\;
		\If {$OF(s) \geq v$}{
			output $s$\;
		}
	}
\end{algorithm}

After acquiring all the rare high-utility sequential rules, the \textit{SWF} and $OF$ of each sequence in the sequential database are calculated according to the above definitions, and details are exhibited in Algorithm 1. The deviation factor $DF(r)$ is calculated initially for each sequential rule $r$. $S(s)$ and $A(s)$ as temporary variables respectively record the number of \textit{UOSRs} and the sum of the deviations of \textit{UOSRs} present in each sequence $s$. Subsequently, the sequence weighting factor \textit{SWF}($s$) and outlier factor $OF(s)$ of the sequence $s$ are calculated according to Definition \ref{def_9} and Definition \ref{def_11}. Assuming the derived $OF(r)$ is greater than $v$, then it is an outlier.

Note that we have obtained the \textit{RHUSRs} in Table \ref{table:result} in accordance with the constraints established previously. It is easy to calculate that $DF(\{a, b, c\} \Rightarrow \{g\})$ = (1 - 0.75)/3 $\approx$ 0.083. Similarly, $DF(\{a, c\} \Rightarrow \{g\})$ $\approx$ 0.083 and  $DF(\{b, c\} \Rightarrow \{g\})$ $\approx$ 0.083. Since these three sequential rules are all existed in sequence $S_1$, we can obtain that $S(S_1)$ = 3 and $A(S_1)$ = 0.249. Thus, the calculation process of \textit{SWF}($S_1$) is 3/3 = 1 and $OF(S_1)$ is 1 - 1 * 0.249 = 0.751. If the supplied $v$ is 0.7, then $S_1$ is an outlier.

\subsection{The Proposed DUOS Algorithm}

In this subsection, we describe the specific process of rule mining in detail. Algorithm 2 demonstrates the complete pseudo-code of DUOS. This algorithm takes a sequential database \textit{SD}, a minimum utility threshold \textit{minutil}, a minimum confidence threshold \textit{minconf}, a maximum support threshold \textit{maxsup}, a minimum support threshold \textit{minsup}, and an outlier threshold $v$ as input. Note that its main goal is to identify high-utility outliers in the sequence database.

\begin{algorithm}[h]
	\label{DUOS}
	\caption{The DUOS algorithm}
	\LinesNumbered 
	\KwIn{\textit{SD}, a sequence database; \textit{minutil}, the minimum utility threshold; \textit{minconf}, the minimum confidence threshold; \textit{minsup}, the minimum support threshold; \textit{maxsup}, the maximum support threshold; $v$, the outlier threshold.} 
	\KwOut{a complete set of \textit{UOSRs}.} 
	scan \textit{SD} once to calculate $SEU(i)$ and $sup(i)$ for each item $i \in I$\;
	$I^* \leftarrow \{i| i \in I$ $\wedge$ $SEU(i) \geq$ \textit{minutil} $\wedge $ $sup(i) \geq $ \textit{minsup}\}\;
	remove item $i$ from \textit{SD} such that $i \notin I^*$ to obtain $SD^*$\;
	scan $SD^*$ once to calculate a bit vector \textit{sids(i)} for each item $i \in I^*$\;
	scan $SD^*$ once again to calculate $SEU(r)$, \textit{sids}($r$), and \textit{sup}($r$) for each rule $r$ whose size is 1 * 1\;
	$R^* \leftarrow \{r|  SEU(r ) \geq$ \textit{minutil} $\wedge$ $sup(r) \geq $ \textit{minsup}\}\;
	construct initially utility tables $UT(r)$ and $RCM(i, j)$ for each rule $r: i \Rightarrow j \in R^*$\;
	
	\For {each rule $r \in R^*$ }{
		calculate $u(r)$ by scanning $UT(r)$, and calculate $conf(r)$ = $sids(r)$/$sids$ (\textit{r.antecedent})\;
		\If {$u(r) \geq$ \textit{minutil} $\wedge$ $conf(r) \geq $ \textit{minconf}  $\wedge$ $sup(r) <$ \textit{maxsup}}{
			\textit{RHUSR} = \textit{RHUSR} $\cup$ $r$\;
		}
		\If {($UT(r)$.\textit{iutil} + $UT(r)$.\textit{lutil} + $UT(r)$.\textit{rutil} + $UT(r)$. \textit{lrutil}) $\geq $ \textit{minutil}}{
			call \textbf{RightExpansion}($r$, \textit{SD}, \textit{minutil}, \textit{minconf}, \textit{minsup}, \textit{maxsup})\;
		}
		\If {($UT(r)$.\textit{iutil} + $UT(r)$.\textit{lutil} + $UT(r)$.\textit{lrutil}) $\geq$ \textit{minutil}}{
			call \textbf{LeftExpansion}($r$, \textit{SD}, \textit{minutil}, \textit{minconf}, \textit{minsup}, \textit{maxsup})\;
		}
	}
	call \textbf{OutlierDetection}(\textit{RHUSRs}, $v$, \textit{maxsup});
\end{algorithm}


DUOS first accesses the sequential database once to calculate the sequence estimated utility and support for each item. Then it filters out some unpromising items according to the Strategy \ref{stra_1}, and puts those promising items into $I^*$. Then it removes these items in the sequential database to obtain the revised database $SD^*$. Next, the algorithm visits the revised database $R^*$ again to construct bit vectors for each items to prepare for later calculation of confidence. Afterwards, another database scan is needed to calculate the sequence estimated utility $SEU(r)$ and support $sup(r)$ of rule $r$, as well as the set of sequence that supports $r$. 

Those rules satisfying the requirements are put into $R^*$ and  made as initial rules of 1*1. Thereafter, the utility tables $UT(r)$ of rule $r$: $X \Rightarrow Y$ are constructed, as well as the RCM structure. For each rule $r$ in $R^*$, DUOS supposes it satisfies that $u(r) \geq$ \textit{minutil}, $conf(r) \geq$ \textit{minconf}, and $sup(r) <$ \textit{maxsup} simultaneously, then DUOS puts it into the collection \textit{RHUSR}.  Otherwise, DUOS will determine if it meets the constraints of the Strategy \ref{stra_6}, it expands right; if it satisfies the Strategy \ref{stra_7}, and then it expands left.

\begin{algorithm}[h]
	\label{Right}
	\caption{RightExpansion($r$, \textit{SD}, \textit{minutil}, \textit{minconf}, \textit{minsup}, \textit{maxsup})}
	\LinesNumbered 
	\KwIn{$r$: $X \Rightarrow Y$, a sequence rule; \textit{SD};  \textit{minutil}, the minimum utility; \textit{minconf}, the minimum confidence; \textit{minsup}, the minimum support threshold; \textit{maxsup}, the maximum support threshold.}
	\textit{candidates} $\leftarrow \emptyset$\;
	
	\For {sequence $s \in sids(r)$}{
		\For {each rule $e$: $X \Rightarrow Y$  $\cup$ $ \{i\}|i\in $ \textit{leftRight}($r, s)$ $\cup $ \textit{onlyRight}($r, s) $ $\cup$ \textit{onlyLeft}($r, s)$}{
			\If {$RCM(j, i) \geq $ \textit{minsup}, $j \in X$ }{
				\textit{candidates} $ \leftarrow $ \textit{candidates} $\cup$ $e$\;
				update $UL(e)$\;
			}
		}
	}
	
	\For {each rule $r \in $ \textit{candidates}}{
		\If {$sup(r)$ $<$ \textit{minsup}}{
			continue\;
		}
		\If {($u(r)$ $\geq $ \textit{minutil} $\wedge$ \textit{conf(r)} $ \geq$ \textit{minconf} $\wedge$ \textit{sup(r)} $<$ \textit{maxsup})}{
			\textit{RHUSR} = \textit{RHUSR} $\cup$ $r$\;
		}
		\If {($UT(r)$.\textit{iutil} + $UT(r)$.\textit{lutil} + $UT(r)$.\textit{rutil} + $UT(r)$. \textit{lrutil}) $\geq $ \textit{minutil}}{
			call \textbf{RightExpansion}($r$, \textit{SD}, \textit{minutil}, \textit{minconf}, \textit{minsup}, \textit{maxsup})\;
		}
		\If {($UT(r)$.\textit{iutil} + $UT(r)$.\textit{lutil} + $UT(r)$.\textit{lrutil}) $\geq$ \textit{minutil}}{
			call \textbf{LeftExpansion}($r$, \textit{SD}, \textit{minutil}, \textit{minconf}, \textit{minsup}, \textit{maxsup})\;
		}
	}
\end{algorithm}

Algorithm 3 shows the pseudo-code for the right extension. DUOS finds out the sequential rules that can be expanded at first. In this case, DUOS utilizes the two designed strategies, Strategy \ref{stra_4} and Strategy \ref{stra_5} (Lines 4--7) and takes advantage of the support of shorter rules to reduce the number of candidate rules. Considering Strategy \ref{stra_3}, suppose the support  of a candidate rule $r$ is less than \textit{minsup}, then this rule is definitely not a rare high-utility sequential rule (Lines 11--13). Therefore, there is no need to operate down. It is advisable to jump out of the loop and determine the next sequence rule. Next, DUOS determines whether the rule is qualified or expandable, just like Algorithm 2. As for the left expansion exhibited in Algorithm 4, it is similar to the right expansion except that there is no right expansion operation after the left expansion.

\begin{algorithm}[h]
	\label{Left}
	\caption{LeftExpansion($r$, \textit{SD},\textit{minutil}, \textit{minconf})}
	\LinesNumbered  
	\KwIn{$r$: $X \Rightarrow Y$, a sequence rule; \textit{SD};  \textit{minutil}, the minimum utility; \textit{minconf}, the minimum confidence.} 
	\textit{candidates} $\leftarrow \emptyset$\;
	\For {sequence $s \in$ \textit{sids(r)}}{
		\For {each rule $e$: $X \cup \{i\}$ $\Rightarrow$ $Y|i\in $ \textit{leftRight}($r, s)$ $\cup$ \textit{onlyLeft}($r, s$)}{
			\If {$RCM(i, j) \geq $ \textit{minsup}, $j \in Y$ }{
				\textit{candidates} $ \leftarrow $ \textit{candidates} $\cup$ $e$\;
				update $UL(e)$\;
			}
		}
	}
	\For {each rule $r \in$ \textit{candidates}}{
		\If {$sup(r)$ $<$ \textit{minsup}}{
			continue\;
		}
		\If {$u(r) \geq$ \textit{minutil} $\wedge$  \textit{conf(r)} $\geq $ \textit{minconf} $\wedge$ \textit{sup(r)} $<$ \textit{maxsup}}{
			\textit{RHUSR} = \textit{RHUSR} $\cup$ $r$\;
		}
		\If {($UT(r)$.\textit{iutil} + $UT(r)$.\textit{lutil} + $UT(r)$.\textit{lrutil}) $\geq$ \textit{minutil} and \textit{conf(r)} $ \geq $ \textit{minconf}}{
			call \textbf{LeftExpansion}($r$, \textit{SD}, \textit{minutil}, \textit{minconf}, \textit{minsup}, \textit{maxsup})\;
		}
	}
	
\end{algorithm}

\section{Experiments} \label{sec:experiments}

In this section, we conduct several experiments (real-word datasets) to demonstrate the effectiveness and efficiency of our proposed DUOS algorithm according to rare high-utility rules to detect outlier sequential patterns. 

\textbf{Evaluation metric}. In general, the comparison of evaluation metric for rare high-utility sequential rule mining algorithms consists of efficiency analysis with running time and memory consumption, and scalability evaluation. And then we discuss the effectiveness of outlier detection from rare high-utility sequential rules. In the following subsections, we will take these metrics into account to evaluate the performance of our novel algorithm. 

\textbf{Compared baseline}. DUOS is the first algorithm which aims to detect utility-aware outlier sequential rules from sequence data. For efficiency analysis, we select the HUSRM algorithm   as the baseline. In order to evaluate the effect of the proposed different pruning strategies, three variants of DUOS (respectively denoted as DUOS$_{p1}$, DUOS$_{p2}$ and DUOS) are compared. DUOS is a hybrid optimization algorithm as we introduced before. And the differences between DUOS$_{p1}$, DUOS$_{p2}$ and DUOS are that DUOS$_{p1}$ does not remove those low utility or low support items before construct rules (refers to Strategy \ref{stra_1}). DUOS$_{p2}$ adopts Strategy \ref{stra_1} instead of pruning low utility or low support 1-rules (refers to Strategy \ref{stra_2}) during mining process. And DUOS is the complete algorithm which achieve all effective strategies. Moreover, the effectiveness of different structures is also considered in our experiment, and it tests which structure (array-list and bit-vector) is more efficient to compute confidence of super-rules. The DUOS$_{\textit{list}}$ means algorithm applies array-list to record information about sub-rules, and DUOS$_{\textit{vector}}$ utilizes bit-vector to calculate confidence of super-rules. Finally, we will demonstrate the effectiveness of DUOS in outlier detection by discovering rare high-utility sequential rules.

\subsection{Data Description and Experimental Setup}

\textbf{Datasets}. In the performance experiments, a whole of four datasets were chosen for the different features they displayed. We aim to show the efficiency of the novel algorithm in a wide range of situations for detecting anomaly (outlier patterns). The chosen datasets and their characteristics are listed in Table \ref{table:Dataset}. Here are some feature labels of selected datasets: \#$|D|$ is the amount of sequences;  \#$|I|$ is the number of distinct items in the dataset;  \#S is the length of a sequence $s$; and \#Seq is the number of elements per sequence. Four datasets, including Bible, BMS, Kosarak, and Sign, are used in the experiments. In each dataset, notice that the internal utility of each item and the quantities of items in every transaction are generated using a simulation model \cite{tseng2012efficient,gan2020proum}. All the datasets can be download from SPMF open-source website\footnote{http://www.philippe-fournier-viger.com/spmf/}. In order to make the mining result more reliable, we use support rate times the maximum support of single item in its dataset to set their frequency values. The details can be seen in Table \ref{table:SuppRate}. Note that the rare patterns not only have low-support, but also may high-utility, so that we set two different ranges in Bible (min: 0.3\% and max: 0.9\%) and Sign (min: 44\% and max: 75\%), respectively.

\begin{table}[!htbp]
	\centering
	\caption{Dataset characters}
	\label{table:Dataset}
	
	\begin{tabular}{|c| c | c | c | c | c |}
		\hline
		 \textbf{Dataset} & \textbf{\#$|D|$} & \textbf{\#$|I|$} & \textbf{avg (\#S)} & \textbf{max (\#S)} & \textbf{avg (\#Seq)}  \\
		\hline \hline

	 	Bible  & 36,369  & 13,905 & 21.64 & 100 & 17.85 \\
		BMS  & 77,512  & 3,340  & 4.62  & 267 & 2.5 \\					
		 Kosarak10k & 10,000  & 10,094 & 8.14  & 608 & 8.14   \\
		Sign  	 & 730     & 267    & 52 	& 94  &  51.99   \\

		\hline
	\end{tabular}
\end{table}

\textbf{Experimental setup}. All the compared algorithms in the experiments were implemented in Java language. Note that the original HUSRM algorithm ignores influence of the frequency metric in the final results and outlier detection. Thus, the output of HUSRM used here may cause some differences. Besides, we firstly take attention of utility factor when comparing HUSRM with different variants of the DUOS algorithm. Then we present an analysis about anomaly detection effectiveness of DUOS. All experiments are performed on a workstation with an Intel(R) Core(TM) i7-9700 CPU @ 3.00 GHz 3.00 GHz, 16 GB assigned RAM, and with the 64-bit Microsoft Windows 10 Home Edition operating system.

\begin{table*}[!htbp]
	\centering
	\small
	\caption{Experiments on Sign dataset}
	\label{table:SignResult}
	\begin{tabular}{|c|c|llllllll|}
		\hline\hline
		\textbf{Sign} & \textbf{\textit{minutil} (\$)}  & 28k & 30k & 32k & 34k & 36k & 38k & 40k & 42k \\ \hline
		
		&\textbf{HUSRM} 		  &  -   &   -  & 2127.3 & 1766.6 & 1479.2 & 1125.5 & 963.7 & 764.9 \\
		\textbf{Runtime (s)} & \textbf{DUOS$_{p1}$}  & 14.9 & 14.9 & 15   & 13.7 & 14.1 & 13.9 & 12.8 & 12.3 \\
		&\textbf{DUOS$_{p2}$}  & 15.5 & 13.5 & 13.8 & 14.1 & 12.1 & 11.8 & 12.4 & 12.5 \\
		&\textbf{DUOS}   	  & 14.3 & 13.5 & 13.3 & 13   & 12.9 & 11.7 & 11.6 & 11 \\ \hline
		
		&\textbf{HUSRM} 		  &   -   &   -   & 536.1 & 535.4  & 533.1 & 533.9 & 535.9 & 532.7 \\
		\textbf{Memory (MB)}&\textbf{DUOS$_{p1}$}  & 541.9 & 542.6 & 540.9 & 539.78 & 535.7 & 535.1 & 534.1 & 533.4 \\
		&\textbf{DUOS$_{p2}$}  & 544.1 & 542.8 & 539.6 & 537.2  & 537.7 & 538.3 & 536.7 & 537.7 \\
		&\textbf{DUOS}   	  & 541.1 & 540.7 & 538.3 & 535    & 533.8 & 532   & 530.5 & 530.1 \\ \hline
		
		&\textbf{HUSRM} 		 &  -  &  -  & 332 & 181 & 91 & 43 & 18 & 7 \\
		\textbf{Rules}&\textbf{DUOS$_{p1}$}  & 287 & 214 & 156 & 108 & 67 & 37 & 16 & 7 \\
		&\textbf{DUOS$_{p2}$}  & 287 & 214 & 156 & 108 & 67 & 37 & 16 & 7 \\
		&\textbf{DUOS}   	     & 287 & 214 & 156 & 108 & 67 & 37 & 16 & 7 \\ 
		
		\hline\hline
	\end{tabular}
\end{table*}

\begin{table*}[!htbp]	
	\centering
	\caption{Final rules and outliers on each dataset}
	\label{table:Outlier}

	\begin{tabular}{|c|c|llllllll|}
		\hline\hline
		\multicolumn{2}{|c}{\textbf{Dataset}} & \multicolumn{8}{|c|}{\textbf{\# \textit{minConf} = 0.7 and \textit{minOutlier} = 0.9}}\\ \hline
		
		& \textbf{\textit{minutil} (\$)} & 5k   & 6k   & 7k   & 8k   & 9k   & 10k  & 11k & 12k \\ \cline{2-10}
		\textbf{Bible}&\textbf{rule} 		   		 & 257  & 229  & 196  & 161  & 129  & 99   & 61  & 36  \\
		&\textbf{outlier}  			 & 1834 & 1759 & 1526 & 1413 & 1230 & 1211 & 916 & 668 \\ \hline
		
		& \textbf{\textit{minutil} (\$)} & 4k   & 5k   & 6k   & 7k   & 8k   & 9k  & 19k & 11k \\ \cline{2-10}
		\textbf{BMS}&\textbf{rule} 		       & 28   & 27   & 22   & 14   & 12   & 8   & 5   & 3   \\
		&\textbf{outlier}  		   & 1630 & 1630 & 1538 & 1239 & 1172 & 867 & 695 & 556 \\ \hline
		
		& \textbf{\textit{minutil} (\$)} & 6k   & 8k  & 10k & 12k & 14k & 16k & -  & - \\ \cline{2-10}						 
		\textbf{Kosarak10k}&\textbf{rule} 		   		  & 20   & 16  & 14  & 8   & 7   & 5   & - & - \\
		&\textbf{outlier}  			  & 1095 & 888 & 867 & 531 & 419 & 289 & - & - \\ \hline
		
		& \textbf{\textit{minutil} (\$)}	 & 28k & 30k & 32k & 34k & 36k & 38k & 40k & 42k \\ \cline{2-10}	 
		\textbf{Sign}&\textbf{rule} 			 	 & 287 & 214 & 156 & 108 & 67  &  37 & 16  & 7   \\
		&\textbf{outlier}  		 	 & 717 & 709 & 661 & 636 & 602 &  575& 568 & 559 \\
		
		\hline\hline
	\end{tabular}
\end{table*}

\begin{table}[!htbp]	
	\centering
	\small
	\caption{Support rate on each dataset}
	\label{table:SuppRate}
	\begin{tabular}{|c|c|c|c|}
		\hline\hline
		\textbf{Dataset} & \textbf{\textit{\textit{maxsup}}} & \textbf{Minimum rate} & \textbf{Maximum rate} \\ \hline
		
		Bible & 26577 & 0.3\% (80) & 0.9\% (239) \\ \hline
		
		BMS & 3658 & 3\% (110) & 9\% (329) \\ \hline
		
		Kosarak10k & 6058 & 2\% (121) & 5\% (302) \\ \hline
		
		Sign & 680 & 44\% (299) & 75\% (510) \\ 
		
		\hline\hline 
	\end{tabular}
\end{table}

\subsection{Efficiency Analytic}

\begin{figure*}[htbp]
	\centering
	\includegraphics[trim=130 0 10 0,clip,height=0.15\textheight,width=1.1\textwidth,scale=0.5]{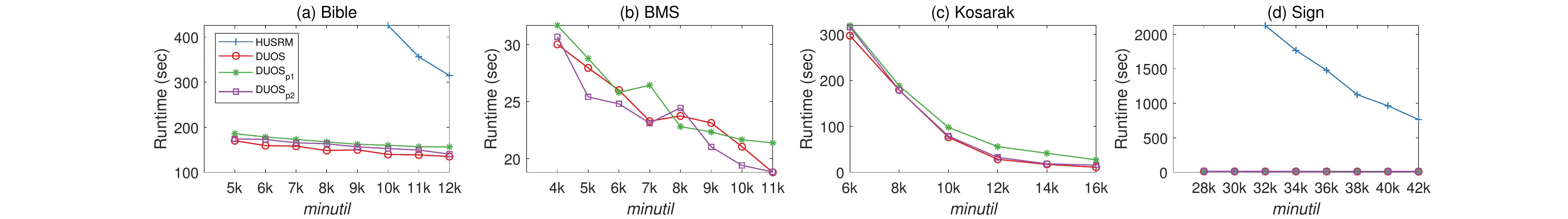}
	\captionsetup{justification=centering}
	\caption{Runtime by varying \textit{minutil}.}
	\label{fig:Runtime}	
\end{figure*}

As is known to all, an acceptable sequence mining algorithm should be efficient enough and scale-well to handle massive datasets. Therefore, we evaluate the execution time of the compared methods under different parameter settings. In each distinct dataset, we increase the \textit{minutil} threshold and do not change other parameters. However, with the different features in those tested datasets, we may not set the same values to all parameters. All the tested data size is fixed, and we finally obtain the running time of the results in Fig. \ref{fig:Runtime} under each parameter setting. Notice that if the experimental algorithm's runtime spends more than 10,000 seconds, we suppose this algorithm is unacceptable or useless, and we do not show here.

It is interesting to observe that DUOS always runs deeply faster than HUSRM. For example, when conducted on BMS and Kosarak10k, it is clearly that HUSRM has always run out of exceed time though we test the biggest \textit{minutil} value (\$11,000 and \$16,000). Consider the Bible as shown in Fig. \ref{fig:Runtime}(a), among the four algorithms, HUSRM is the most time-consuming: when \textit{minutil} is setting from \$5,000 to \$9,000, it does not output results within the time limitation (w.r.t. 10,000 seconds). When \textit{minutil} = \$12,000, it still takes the runtime as 314 seconds, which is quite longer than other algorithms. The performance of DUOS, DUOS$_{p1}$, and DUOS$_{p2}$ respectively keeps a slow downward trend, and starts to nearly maintain stable afterwards. While in another dataset Sign, HUSRM performs worse than DUOS and its variants. And its average run time is over one orders of magnitude than that of DUOS. In order to make Fig. \ref{fig:Runtime}(d) result easier to observe, we list the details of testing Sign dataset in Table \ref{table:SignResult}. Firstly, the amount of final high-utility rules explains why the runtime gap is huge between DUOS and HUSRM. For instance, while \textit{minutil} is \$32,000, DUOS algorithm outputs 176 rules less than HUSRM. Although they have same \textit{minutil} threshold, DUOS considers the rare patterns whose supports belong to an interval range. The frequency limit makes DUOS prune many ineligible items at the beginning and unqualified rules during the mining process.

\begin{figure}[!htbp]
	\centering
	\includegraphics[trim=70 30 0 0,clip,height=0.33\textheight,width=1.1\columnwidth]{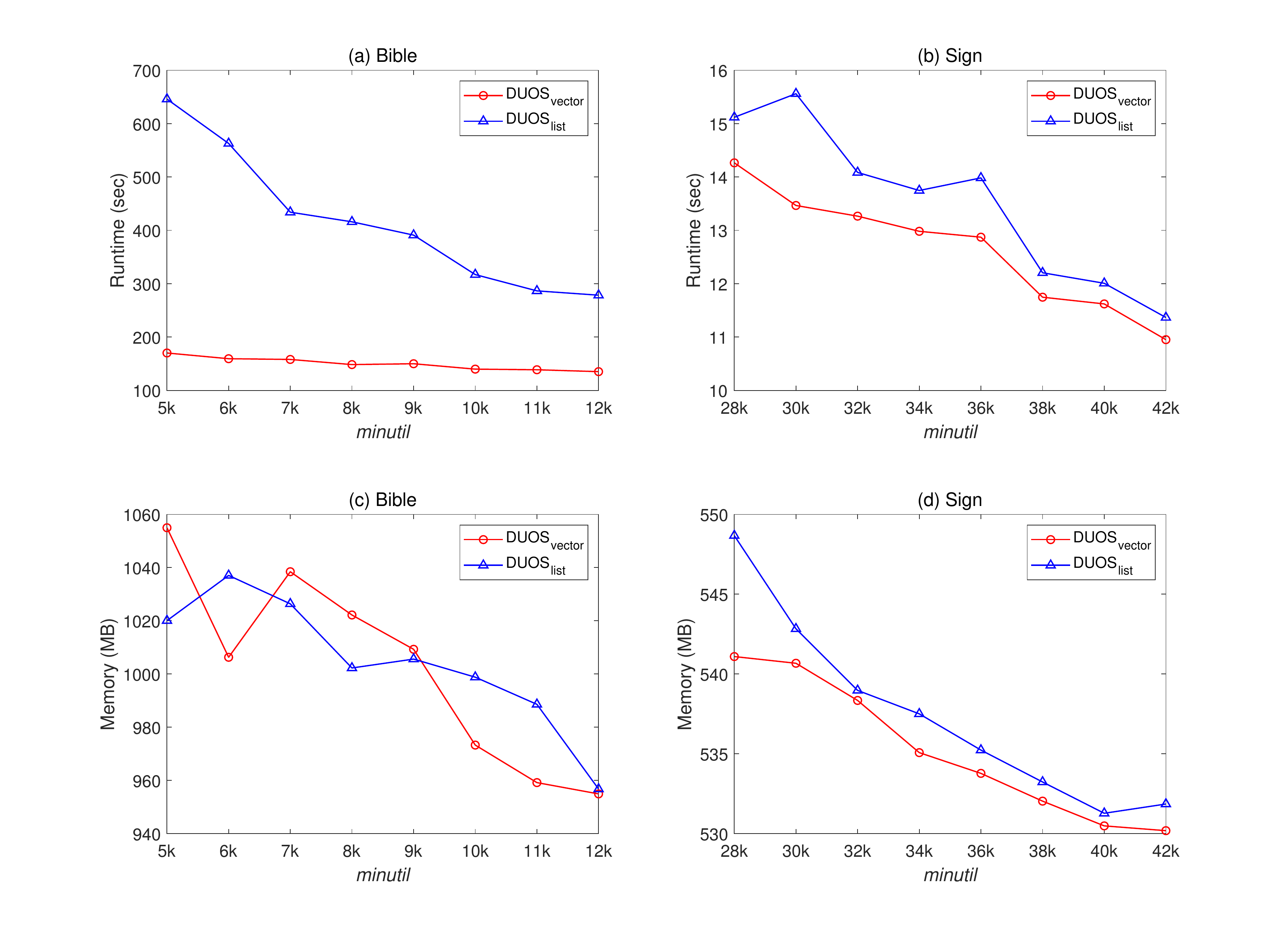}
	\captionsetup{justification=centering}
	\caption{Comparison between array-list and bit-vector.} 
	\label{fig:structs}
\end{figure}

In Fig. \ref{fig:structs}, intuitively, we can observe that to utilize bit-vector structure is more efficient than adopt array-list to compute the confidence of super-rules from each sub-figure. We select two distinct datasets, which Bible has 13,905 unique items and Sign only owns 267 items. As we introduced before, DUOS$_{\textit{vector}}$ algorithm uses ``Key-Values'' to store information of sub-rules, thus it just needs to find out the corresponding ``Keys'' when calculating the super-rules. On the contrary, DUOS$_{\textit{list}}$ has to access array-list to compose super-rules and this is too time-consuming. That's why DUOS$_{\textit{list}}$ always spend more time from Fig. \ref{fig:structs}(a) and Fig. \ref{fig:structs}(b). Furthermore, we also can observe experimental algorithms has slight gap in Fig. \ref{fig:structs}(b) but not in Fig. \ref{fig:structs}(a). For example, when \textit{minutil} is \$5,000 on Bible dataset, the runtime of DUOS$_{\textit{list}}$ is close to 650 seconds, while DUOS$_{\textit{vector}}$ approximately takes 180 seconds. Notice that the biggest gap on Sign dataset is around 2 seconds in \textit{minutil} = \$30,000. The reason is that Bible has 36,369 sequences which is bigger than Sign, while the \textit{minutil} threshold in Sign is from \$28,000 to \$42,000, which is nearly three times than Bible experiment sets. Among the memory consumption comparisons, following the \textit{minutil} increases, the gap between DUOS$_{\textit{vector}}$ and DUOS$_{\textit{list}}$ is more and more narrow. It is because the bigger threshold sets, the less suitable rules be discovered. If we give a large value as  the initial \textit{minutil}, many items will be pruned directly at the beginning.

\textbf{Summary of efficiency analytic}. The above results demonstrate the efficiency of DUOS. Under different parameter settings (when \textit{minutil} is large), DUOS always takes less time than the existing HUSPM algorithm.  In addition, the bit-vector structure is applied to confidence computing procedure during the recursive mining processes. All these experimental results demonstrate that the proposed DUOS model has a suitability for dealing with real datasets.

\subsection{Memory Evaluation}

\begin{figure*}[htbp]
	\centering 
	\includegraphics[trim=120 0 10 0,clip,height=0.15\textheight,width=1.1\textwidth,scale=0.5]{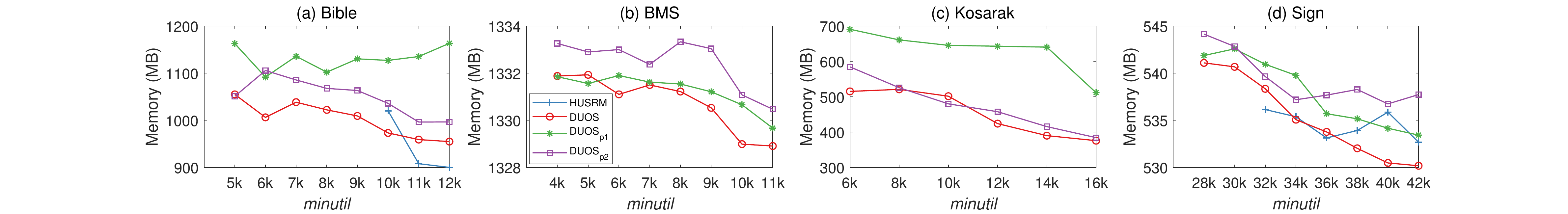}
	\captionsetup{justification=centering}
	\caption{Memory usage by varying \textit{minutil}.}
	\label{fig:Memory}	
\end{figure*}

In this subsection, we further evaluate the mining efficiency of DUOS in terms of memory usage. Unless otherwise stated, all parameters are set to the default values as that in Fig. \ref{fig:Runtime}.  Results of the peak memory consumption of all the compared algorithms are plotted in Fig. \ref{fig:Memory}(a) to Fig. \ref{fig:Memory}(d), respectively. Note that we use the Java API to calculate the peak memory usage of each algorithm during the whole mining process.

As we can see, HUSRM seems to perform significantly better than DUOS and its variants with large \textit{minutil} values from Bible and Sign datasets. However, when \textit{minutil} sets \$9000, the runtime cost over 1,000 seconds (in order to make sub-figure more suitable, we do not show the point). And HUSRM always runs overtime in BMS and Kosarak datasets, so that we can deduce it is still the worst algorithm in our tests. From all sub-figures in Fig. \ref{fig:Memory}, we can easily observe that DUOS performs significantly better than other variants model. The reason have been analyzed in previous efficiency analytic subsection. 

In addition, the more interesting things is that DUOS$_{p1}$ performs not very well compares to DUOS$_{p2}$. Specially, it is strongly visible in Fig. \ref{fig:Memory}(c), when \textit{minutil} = \$14,000, DUOS$_{p1}$ almost consumes 250 MB more than DUOS$_{p2}$ needs. On the contrary, as shown in Fig. \ref{fig:Memory}(b), we also can find DUOS$_{p2}$ works worse than DUOS$_{p1}$. For example, the peak memory consumption for DUOS$_{p1}$ is significantly less than that of DUOS$_{p2}$ (\textit{minutil} = \$6000). The reason causes this total different results is the features of dataset. DUOS$_{p1}$ aims to prune those low-support or low-utility items before generate rules. The number of candidates has been reduced from the start. What's more, DUOS$_{p2}$ deletes unpromising 1-rules before constructs high level rules during the mining process. Table \ref{table:Dataset} shows that the average length of sequence of BMS is shortest among the experimental datasets, which may cause that almost items may not need to be removed at the beginning. And the item pruning strategy may not work well here and it still generates many unpromising rules later.

As shown on all datasets, the memory usage of DUOS, DUOS$_{p1}$ and DUOS$_{p2}$ has obvious changes. From the Table \ref{table:SignResult}, the last row shows the change of high-utility rules generated following the increased \textit{minutil}. As we introduced before, the higher \textit{minutil} we set, the less promising rules will be discovered. Then the runtime and memory consumption will decrease too. In addition, a very large \textit{minutil} brings no extra benefit to the discovered results. Hence, these thresholds should not be set too large in practice.

\textbf{Summary}. The proposed DUOS model with several pruning strategies consumes less memory than HUSRM under most parameter settings. DUOS has the least memory consumption on Fig. \ref{fig:Memory}(a) to (d). Nonetheless, for these cases, the best performing DUOS on memory consumption has much worse execution time in Fig. \ref{fig:Runtime}(b). As mentioned previously, one of the advantages of DUOS is that it is able to early filter a large amount of frequent patterns and avoid generating some unpromising rules in later processes.

\subsection{Scalability Test}

\begin{figure}[!htbp]
	\centering
	\includegraphics[trim=70 30 0 0,clip,scale=1.2,height=0.33\textheight,width=1.12\columnwidth]{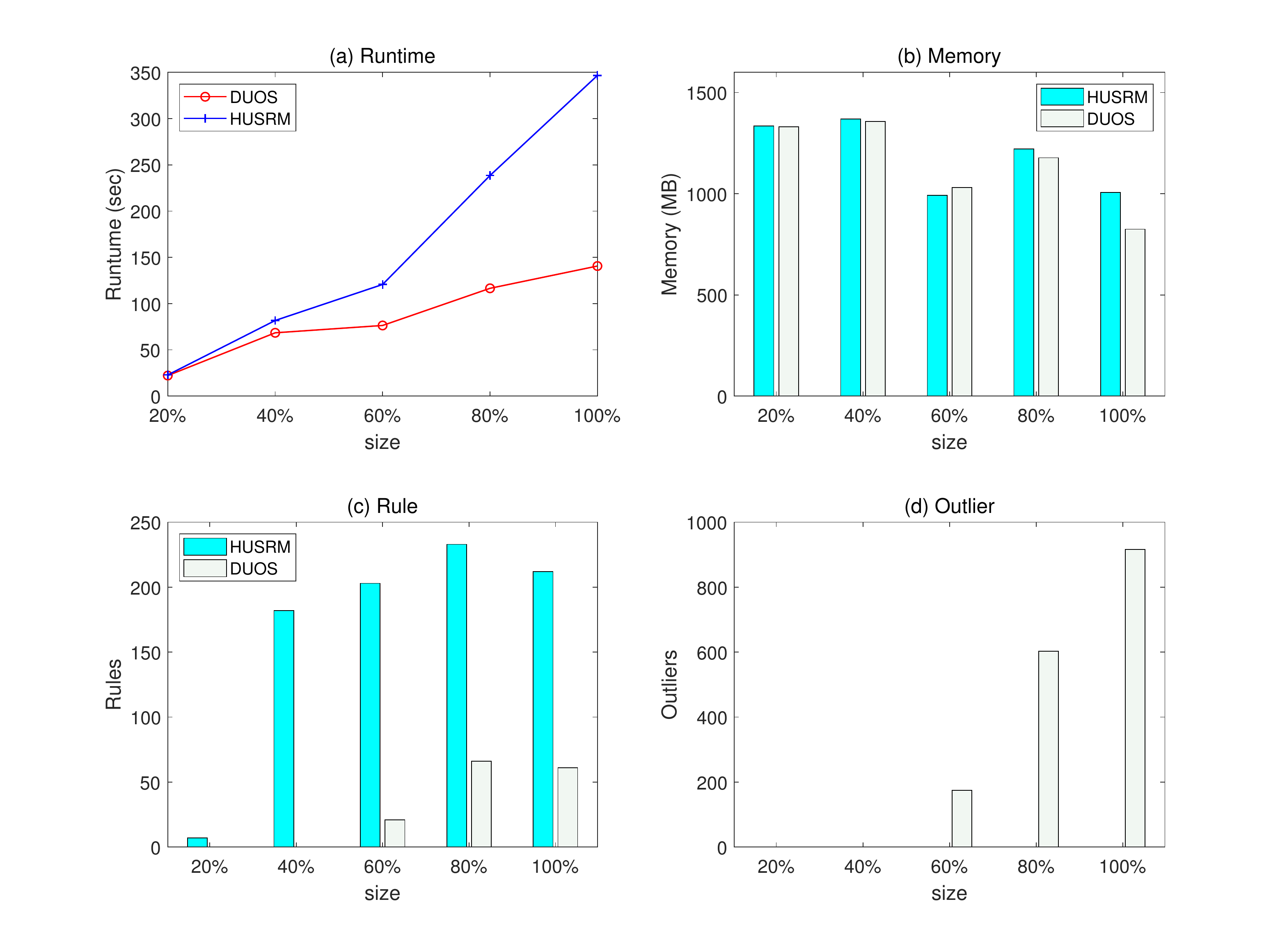}
	\captionsetup{justification=centering}
	\caption{Scalability test.}
	\label{fig:Scalability}
\end{figure}

In real life, the data (especially the large sequence data) is always so massive that scalability of an algorithm is vital so much. Thus, in this subsection, we compare DUOS and HUSRM analytically on Bible dataset. The experimental results are shown in Fig. \ref{fig:Scalability}.

We mainly test Bible dataset in terms of different data sizes: from 7k (20\%) to 36k (100\%) sequences. We adopt the same parameters as previous settings, the \textit{minSupRate}, \textit{maxSupRate} and \textit{minConf} are 0.3\%, 0.9\%, and 70\%, respectively. Due to the HUSRM algorithm does not consider outlier patterns, we set a default value 0 as its output, and \textit{minOutlier} was set as 90\% in our proposal DUOS algorithm. We will analyze outlier generation performance in next subsection.

Fig. \ref{fig:Scalability}(a) shows the result of runtime performance in terms of increasing data sizes. The DUOS algorithm shows superior scalability that time consumption grows linearly and steadily. Besides, an interesting observation can be seen in Fig. \ref{fig:Scalability}(b), both HUSRM and DUOS reduce the memory usage when data size increase. The reason is that the values of items in Bible are small and most of them are concentrated in the head part, then they have to cost a lot of memory to store information. For example, when data size is 20\%, DUOS and HUSRM both consume nearly the same amount of memory (1330 MB). However, their cost decreases significantly in 100\% data size. According to the generated high-utility sequential rules, as shown in Fig. \ref{fig:Scalability}(c), these results are within our expectation. Clearly, the rare support metric can reduce the amount of final rules.

\textbf{Discussion}. In summary, the scalability experiments confirm the intuition that DUOS using bit-vector structure and pruning strategies is widely scalable for a large-scale dataset, and is superior to the existing algorithms.

\subsection{Outlier Detection Analytic}

Generally, if we use positive detection, it may make sense to remove outliers (extreme values), that place an upper-bound on the number of deviations from the user-specified standard deviation for outliers. On the contrary, if negative detection is used, it also makes sense to keep the outliers and separate them from the rest of the data (possibly in a separate container), since the presence of outliers usually indicates contextual anomalies \cite{cao2019efficient}. Our proposal algorithm utilizes sequential rules to detect anomaly automatically. It detects these rare high-utility rules in advance, if a sequence contains many these rules, this sequence more likely is an anomaly (outlier) sequence. Table \ref{table:Outlier} shows the quantitative relationship between rules and outlier sequences. The ``rule''  means the number of rare high-utility rules discovered by DUOS, and ``outlier'' represents the quantity of sequences which contain those rules in datasets.  It should be noted that we only test six points on Kosarak10k dataset, and others we fill with ``-'' symbol. If we look carefully, we can observe that when \textit{minutil} = \$16,000, DUOS only found out 5 suitable rules.

When setting \textit{minutil} = \$28,000 in Sign dataset, DUOS finally discovered 287 rules and shown that there are 717 sequences may be outliers. However, the total number of sequence of Sign is actually only 730, many of them are not outliers. In this case, the reason may be the \textit{minutil} value is too low. After that, when setting \textit{minutil} as \$42,000, there still exists 559 sequences admitted as abnormal, although DUOS merely discovered 7 rare high-utility sequential rules. In fact, the items in Sign dataset may be strongly connected, while few of them or even none are outlier patterns.

Consider the BMS dataset, when \textit{minutil} = \$4,000, total 1630 sequences are regarded as anomaly and the number of sequence BMS has is 77,512 (around 2\%). We argue that those sequences are very much outlier sequences. And as \textit{minutil} increases, the amount of rules discovered by these compared algorithms decreases correspondingly (from 2\% to 0.7\%). How much  this \textit{minutil} threshold is suitable for this task is depended on the real circumstance. For example, heart rate is influenced by factors such as physical exercise state, posture (sitting, standing or lying), other diseases, etc.

\textbf{Summary}. The novel DUOS method detects abnormal patterns by considering only vital parameters. There may also exist some additional factors which can affect these parameters. Furthermore, anomaly detection can be combined with human activity identification to promote effective anomaly detection. The integration of sequence-based data mining with outlier detection can provide a good interpret-ability of discovering results in the form of rare high-utility sequential rules. Therefore, additional metrics, such as utility, risk, and confidence, are more suitable to address this PM4AD issue.

\section{Conclusion and Future Work}   \label{sec:conclusion}

In this work, we design a utility-driven outlier detection framework for the effective discovery of abnormal sequential rules from a sequence data. The designed DUOS algorithm introduces a new pattern semantics called utility-aware outlier sequential rule (UOSR). It can effectively solve the problem of existing pattern mining for anomaly detection (PM4AD) methods that tend to miss the high-utility abnormal patterns violating the typical patterns. To compress the useful information from sequence data, the data structure namely utility table is designed. Moreover, we design several novel mining strategies that can naturally capture the contextual information, both frequency and utility, in which an outlier rule occurs. And the confidence measure is adopted in DUOS to efficiently discover the high-utility outlier rules with a high confidence by mining the sequence data without multiple database scans. Detailed experimental evaluation with several real datasets demonstrates the effectiveness of DUOS in capturing outlier rules, and its efficiency in discovering UOSR.

In the future, we expect to continuously improve the efficiency of the proposed algorithm or apply it to deal with dynamic data \cite{gan2018survey}, heterogeneous information network \cite{peng2021lime}, stream data \cite{hackman2019mining,peng2021streaming}, or other complex applications \cite{gan2018privacy}. Due to the advantages of utility factor in evaluating the outliers, how to utilize the existing advances in utility mining \cite{gan2021survey} to address PM4AD is also interesting. 


\ifCLASSOPTIONcompsoc
  \section*{Acknowledgments}
\else
  \section*{Acknowledgment}
\fi

This work was partially supported by National Natural Science Foundation of China (Grant Nos. 62002136 and 61902079), Guangzhou Basic and Applied Basic Research Foundation (Grant Nos. 202102020277 and 202102020928).

\ifCLASSOPTIONcaptionsoff
  \newpage
\fi

\bibliographystyle{IEEEtran}
\bibliography{DUOS.bib}

\end{document}